\documentclass[sigconf]{acmart}
\renewcommand\footnotetextcopyrightpermission[1]{} % removes footnote with conference information in first column

\usepackage{url}
\usepackage{breqn}
\usepackage{subcaption}

\usepackage[ruled,linesnumbered]{algorithm2e}

\newtheorem{Example}{Example}
\AtBeginDocument{%
  \providecommand\BibTeX{{%
    \normalfont B\kern-0.5em{\scshape i\kern-0.25em b}\kern-0.8em\TeX}}}

\setcopyright{none}
\acmDOI{}
\acmISBN{}
\begin{document}

\title{Towards Split Learning-based Privacy-Preserving Record Linkage}

\author{Michail Zervas}
\email{ics20015@uom.edu.gr}
\affiliation{%
  \institution{Dept. of Applied Informatics, University of Macedonia}
  \streetaddress{Egantia 156}
  \city{Thessaloniki}
  \country{Greece}
  \postcode{54636}
}

\author{Alexandros Karakasidis}
\email{a.karakasidis@uom.edu.gr}
\affiliation{%
  \institution{Dept. of Applied Informatics, University of Macedonia}
  \streetaddress{Egantia 156}
  \city{Thessaloniki}
  \country{Greece}
  \postcode{54636}
}

\renewcommand{\shortauthors}{Zervas and Karakasidis}

\begin{abstract}
Split Learning has been recently introduced to facilitate applications where user data privacy is a requirement. However, it has not been thoroughly studied in the context of Privacy-Preserving Record Linkage, a problem in which the same real-world entity should be identified among databases from different dataholders, but without disclosing any additional information. In this paper, we investigate the potentials of Split Learning for Privacy-Preserving Record Matching, by introducing a novel training method through the utilization of Reference Sets, which are publicly available data corpora, showcasing minimal matching impact against a traditional centralized SVM-based technique.
\end{abstract}





\maketitle
\thispagestyle{empty}
\section{Introduction}
The entity resolution problem boils down to identifying the same real world entity in distinct datasets. When these entities are structured in relational records, we refer to the record linkage problem. As these datasets originate from databases of different dataholders, common unique identifiers are not usually available. In this case, combinations of attributes forming a candidate key may be employed. When combined, these attributes, called \textit{quasi-identifiers}, can uniquely identify an entity.

A basic problem with quasi-identifiers is that they are usually the result of manual input. As such, they suffer from problems as misspellings and typos. To this end, in the case of strings, to be able to compare possibly misspelled attributes from different records distance or similarity measures are used \cite{Chr12} with certain thresholds. When a pair of records agrees within these thresholds for each of its corresponding attributes, then this pair of records is considered as matched and these two records are linked.

What has been described so far is the classical version of the record linkage problem. However, in the recent years, the situation has become even more complicated, including a variety of reasons, ranging from emerging privacy-oriented legislation around the world, such as GDPR or HIPAA, business confidentiality in terms of competition, and so on. To this end, the direct plaintext comparison of record attributes becomes problematic as it reveals sensitive or classified information. A typical example of such a case is linking patients records from datasets of different organizations, e.g. in terms of medical research. 

To address these problems, privacy-preserving techniques have emerged where, instead of revealing the entire dataset (e.g. through transferring), only the linkage results are revealed to the dataholders participating in the process. This problem is called Privacy-Preserving Record Linkage (PPRL) and it aims at identifying records describing the same real world entities, i.e. individuals, across diﬀerent databases while not revealing any other information.

In this work, we showcase our efforts for employing a \textit{Split Learning} (SL) \cite{gupta2018distributed} approach for PPRL. The proposed methodology is versatile and able to accommodate a variety of Machine Learning algorithms. To demonstrate its operation, we employ Support Vector Machines, used in the past for the classical version of the Record Linkage problem \cite{christen2008automatic}. Split Learning is a fairly recently introduced technique aimed at training AI models without the need for direct data interchanges. Instead, intermediate representations, coined as smashed data in the SL terminology are used. To utilize this methodology we created these representations employing Reference Sets (RS), i.e. publicly available corpora that can be used as an intermediate point of reference for data obfuscation \cite{pang2009privacy}. 

The proposed technique offers a series of advantages, compared to state-of-the-art approaches \cite{GkoulalasDivanis21}. First, it does not require a third party playing the role of linkage unit. Second, it maintains privacy as, through the use of Reference Sets, no encoded data are transferred, but tensors of distances. Third, it manages to achieve a very low loss in terms of matching accuracy. To this end, in this paper:
\begin{itemize}
    \item We introduce a protocol for Privacy-Preserving Record Matching utilizing SVMs for enabling Split Learning.
    \item We provide a methodology for training individual SVM models at the dataholders without the need of data interchange.
    \item We provide empirical evidence of the performance of our method.
\end{itemize}

The rest of this manuscript is organized as follows. Section \ref{sec:related} contains related work. In Section \ref{sec:background} we formally present the problem we address and provide the necessary prerequisites for this work. We present our method in Section \ref{sec:methodology} and its evaluation in Section \ref{sec:evaluation}. Finally, we conclude and lay out our next steps in Section \ref{sec:conclusions}.
\section{Related Work}
\label{sec:related}

%To perform PPRL, to the best of our knowledge, the majority of approaches follows two directions. In the first one, each dataset is encrypted into some encoding, with Bloom Filters being one of the most common choices, and a third party, usually referred to as the Linkage Unit for matching these encodings. The other direction is employing cryptographic protocols as Homomorphic Encryption. However, in the first case utilizing a third party is not always feasible, while, in the second case, these protocols are, up to now, computationally expensive.

A recent survey on PPRL may be found in \cite{GkoulalasDivanis21}. One of the most popular approaches to this problem regards the use of Bloom filters \cite{Schnell09,ranbaduge2023privacy}, which are combined with \textit{n-grams} and the resulting bit vectors are ANDed for assessing matching status. For this purpose, a separate server referred to as the \textit{Linkage Unit} is used. Our approach does not require such an external entity. Furthermore, such solutions have certain vulnerabilities, requiring additional hardening measures \cite{franke2021evaluation}. 

Beyond Bloom Filters, bit vectors and Locality Sensitive Hashing have been combined \cite{smith2017secure}, at an increased computational cost \cite{christen2020linking}. A two-step hash method with quasi-identifiers converted into \textit{n-grams} is proposed in \cite{ranbaduge2020secure}, requiring, however, a Linkage Unit and a threshold to perform approximate matching. Differential privacy has also been used for PPRL, with current solutions focusing on categorical and numerical attributes \cite{RaoCBK19}, while this work focuses on strings.

Cryptography-based methods have been also proposed, exhibiting, nevertheless, high computational cost. Homomorphic encryption is also susceptible to certain types of attacks\cite{essex2019secure}. Garbled circuits \cite{chen2018perfectly}, need further investigation regarding size and reusability \cite{saleem2018recent}. Fuzzy Vaults \cite{mullaymeri2021using} relying on polynomial reconstruction through interpolation, should also be further investigated in terms of matching performance. 

Finally, the rise of capabilities of Large Language Models (LLMs) provided a new perspective to the classical Entity Resolution problem offering novel approaches (e.g. \cite{li2024leveraging}) with LLMs used to enhance performance. Nevertheless, to the best of our knowledge, there has not been, up to now, an approach that utilizes LLMs and preserving privacy, at the same time.

\section{Prerequisites}
\label{sec:background}
In this section, the problem to be solved is formally described and the required background of the proposed approach is presented.

\subsection{Problem Formulation}
Without harming the general case, two data sources are considered, called Alice ($A$) and Bob ($B$), who respectively hold $r^A$ and $r^B$ records each. We denote as $r_i^A$ and $r_i^B$ the $i$-th record of Alice and Bob, respectively. We represent the $j$-th attribute of these records as $r_i^A.j$ and $r_i^B.j$.

\textit{Privacy-Preserving Record Matching} is the problem of matching all pairs of $r^A$ and $r^B$ records referring to the same real world entity, so that no further information is revealed to $A$, $B$ or any other party involved except for the identifiers of the linked $r^A$s and $r^B$s. 

As Alice and Bob's databases are expected to have different schemas, their records have different attributes and do not share any common candidate keys. Let $R^A$ be Alice's schema and $R^B$ be Bob's schema and let us assume that in these schemas $m$ of the attributes are common between the two sources forming a composite key. These attributes are quasi-identifiers being names, surnames, addresses, birth dates and so on, being unable each on its own to uniquely identify a record. We refer to these attributes as \textit{matching attributes} or \textit{matching fields}. The composite key is used to uniquely identify a record and as such, to determine when two records \textit{match}, i.e., when they refer to the same entity, by comparing the respective attributes. Considering that these data are often dirty, matching should rely on a similarity or distance function. 

Let us consider $\mathcal{D}$ as the domain of each matching attribute, a similarity function $sim_j():\mathcal{D}\times \mathcal{D}\rightarrow[0..1]$ and a threshold $t_j > 0$. Given records $r_i^A$ and $r_i^B$ with matching attributes $r_i.1,\dots, r_i.m$ for both Alice and Bob, we define the following matching function $M:\mathcal{D}\times \mathcal{D}\rightarrow \lbrace 0, 1\rbrace$:
\begin{equation}
	\begin{split}
		M(r_i^A,r_i^B) = \begin{cases} 1, &\mbox{iff } sim_j(r_i^A.j,r_i^B.j) \geq t_j, \forall j \in [1,m] \\ 0, &\mbox{otherwise}. \end{cases}
	\end{split}
	\label{equation:match}
\end{equation}
If $M(r_i^A,r_i^B)=1$, then the pair $(r_i^A,r_i^B)$ is a match.

This formulates \textit{Privacy Preserving Matching} (PPM) and, after this process concludes, the only additional gained information by matching parties should be the identifiers of the matched records.

\subsection{Support Vector Machines}
Support-vector machines (SVMs) \cite{vapnik1999nature} are machine learning models that build optimal separation hyperplanes utilizable in classification problems, by trying to maximize a \textit{margin}, i.e. the distance between this hyperplane and each of the classes to be separated. For this to happen, a labeled training set is required. For non-linearly separable datasets, SVMs use a transformation to a new coordinate space described as the \textit{kernel trick}, where similarity may be computed into the transformed space using the original dataset. This is performed using certain similarity functions called \textit{kernel functions}. The Gaussian RBF kernel function \cite{scholkopf2002learning} is such a function.

\section{Methodology}
\label{sec:methodology}
After presenting our building blocks we now present our approach.
\subsection{Overview}
Let us start by describing the broader image on how to use separately trained SVMs to perform Privacy-Preserving Matching, on the premises of Split Learning.

\begin{algorithm}[t]
	\SetAlgoLined
	\SetKwInput{Input}{Input}
	\SetKwInput{Output}{Output}
 	\SetKwFunction{SelectReferenceSet}{Agree_on_Common_Reference_Set}
        \SetKwFunction{AssociateDataToReferenceSet}{Build_Smashed_Dataset}
        \SetKwFunction{GenerateTrainingData}{Create_Training_Data}
        \SetKwFunction{TrainLocalModels}{Train_Local_Models}
        \SetKwFunction{TrainFederated}{Perform_Split_Training}
        \SetKwFunction{DeliverReferenceRepresentations}{Deliver_Smashed_Data}
        \SetKwFunction{MatchLocally}{Split_Privacy-Preserving Matching}
        \SetKwFunction{ExchangeMatchingRecords}{Exchange_Matching}
     \tcc{Preparation}
	    \SelectReferenceSet()\;
            \AssociateDataToReferenceSet()\;
    \tcc{Training}
        \GenerateTrainingData()\;
        \TrainFederated()\;        
    \tcc{Private Matching}
        \DeliverReferenceRepresentations()\;
        \MatchLocally()\;
        \ExchangeMatchingRecords()\;
	\caption{Protocol Overview.}
	\label{algorithm:protocol}
\end{algorithm}

% \begin{algorithm}[b]
% 	\SetAlgoLined
% 	\SetKwInput{Input}{Input}
% 	\SetKwInput{Output}{Output}
% 	\SetKwInput{r}{- r}
% 	\SetKwInput{RS}{- RS}
% 	\SetKwInput{P}{- P}
% 	\SetKwInput{n}{- n}
%         \SetKwInput{TD}{- TD}
% 	\SetKw{KwBy}{by}
% 	\SetKwFunction{Concatenate}{Concatenate}
% 	\SetKwFunction{Fragment}{Fragment}
% 	\SetKwFunction{toASCII}{to_ASCII}
%         \SetKwFunction{ED}{ED}
% 	\SetKwFunction{Append}{Append}	
% 	\SetKwFunction{SetCoefficient}{SetCoefficient}	
% 	\Input{}                                    
% % 	\SetKwSty{texttt}
% 	\r {A recordset}
% 	\RS {A Reference Set}
% 	\P {A mapping of r to RS}
% 	\Output{}
% 	\TD {A tensor of distances between r and RS}
% % =========
% $TD\gets []$\;
%     \ForEach(\tcp*[h]{for each mapping between r and RS attributes}){$p \in P$}{% 
%         $InnerVector$.\Append(\ED($p$))\;
%     }
%     TD.\Append($InnerVector$)\;
% 	\caption{Real data to Reference Set mapping.}
% 	\label{algorithm:rsmapping}
% \Return TD\;
% \end{algorithm}

\subsection{A Split Learning-based Protocol}
To be able to perform Privacy-Preserving Matching using Split Learning, we devised a protocol which is illustrated in Algorithm \ref{algorithm:protocol}. Initially, the matching parties agree on a common RS and the attributes that will be used (line 1). It is important for the RS to either originate from a totally different domain from the datasets or to be preprocessed so as not to contain data that may be found within the matching datasets. Then, data to be matched at each dataholder are mapped to the RS to create a smashed representation, as illustrated in line 2. Then, each dataholder separately builds its feature vectors to be used for local training (line 3). Line 4 stands for the fourth step which is split training separately taking place at each dataholder. After training, smashed data are exchanged (line 5). Using these representations, each dataholder performs matching individually (line 6). Then, the process concludes with each party delivering matched data identifiers to the other (line 7). The proposed protocol assumes that all communications take place using secure channels.

\subsection{Reference Set \& Data Mapping}
As SVMs are models used for supervised learning, training data are required. In Entity Resolution, and as a result in PPRL, being a special sub-case of this problem, the goal is to identify which pairs of entities are matching and which are not. The same holds for records. In the classical case of record linkage using a SVM \cite{christen2008automatic}, features are built using distances (or similarities) between common fields after a direct comparison. However, in the case of PPRL this cannot occur, as this would comprise a privacy breach. For this purpose, we propose a novel method for generating feature vectors for training the individual SVMs at each dataholder, using their own data. The result is a vector of distances between a dataholder's recordset $r$ and $RS$, the common Reference Set used. 
This process requires as inputs a recordset $r$ and a Reference Set $RS$, both with alphanumeric fields, a mapping $M$ from $r$ to $RS$, common for all matching parties, and produces a distances vector $D$. This mapping will be used both for locally training the SVM and during the matching process. As such, for each record attribute in $r$ a distance or similarity measure is calculated between this record attribute and the attribute of $RS$ participating in the mapping, for all $RS$ records. As more than two candidate pairs of records may have the same distance, instead of relating a matching attribute $r_i$ of $r$ with a single $RS$ attribute, $r_i$ is related with multiple attributes $RS$. As for each record in $RS$ there are $k$ reference attributes, a mapping of $r_i$ can now have up to $k*m*|RS|$ values. We may see details in Example \ref{ex:map_data_to_rs}.

\begin{Example}
\label{ex:map_data_to_rs}
Let us assume that we have the following Reference Set:
[[FirstName: 'CHARLIE', LastName: 'ADLER'],
[FirstName: 'JAY' LastName: 'ADLER']]. We would like to map a record with the following attributes: [FirstName: `ADA', MiddleName: `IVY', LastName: `KING']. Let us also assume that we use Edit Distance \cite{Chr12} as a distance metric. In this case, the following distance calculations will take place:[Ed(ADA, CHARLIE),  Ed(ADA, JAY)] = [6, 3],
[Ed(KING, ADLER),  Ed(KING, ADLER)] = [5, 5],
[Ed(IVY, CHARLIE),  Ed(IVY, JAY)] = [7, 2],
[Ed(IVY, ADLER),  Ed(IVY, ADLER)] = [5, 5]. So the resulting distances vector $D$ will be: $D=$[[6, 3], [5, 5], [7, 2], [5, 5]].
\end{Example}

\begin{algorithm}[t]
	\SetAlgoLined
	\SetKwInput{Input}{Input}
	\SetKwInput{Output}{Output}
	\SetKwInput{r}{- r}
	\SetKwInput{F}{- F}
        \SetKwInput{M}{- M}
	\SetKwInput{P}{- P} 
        \SetKwInput{TD}{- D}
        \SetKwInput{TF}{- TF}
	\SetKw{KwBy}{by}
	\SetKwFunction{MapDataSetToReferenceSet}{MapDataSetToRefSet}	
        \SetKwFunction{Corrupt}{Corrupt}	
        \SetKwFunction{Sim}{d}
        \SetKwFunction{Append}{Append}
        \SetKwFunction{Random}{Random}
	\Input{}                                    
	\TD {Distance vector between r and RS}
        \r {Dataset with alphanumeric fields}
        \M {Mapping between r \& RS}        
	\Output{}
        \F {Array of feature vectors}
$r'\gets$\Corrupt($r$)\;
$D'\gets$ \MapDataSetToReferenceSet($r', RS, M$)\;
$F\gets []$\;
\ForEach(\tcp*[h]{for each row}){$r_i \in r$}{% 
        $F$.\Append([\Sim ($D_i$, $D'_i$), 1])\;
        $k\gets$ $D'_{k\neq i}$\;
        $F$.\Append([\Sim ($D_{i}$, ${D'}_{k}$), 0])\;
    }
\Return F\;    
    \caption{Synthetic Training Data Building.}
    \label{algorithm:create_ts}
\end{algorithm}

\begin{algorithm}[b]
	\SetAlgoLined
	\SetKwInput{Input}{Input}
	\SetKwInput{Output}{Output}
        \SetKwInput{TDA}{- $D^A$}
        \SetKwInput{TDB}{- $D^B$}
        \SetKwInput{MA}{- $MA$}  
        \SetKwInput{SVM}{- $SVM()$}  
	\SetKw{KwBy}{by}
        \SetKwFunction{Sim}{d}
	\Input{}                                    
        \TDA {Distances vector for party A}
        \TDB {Distances vector for party B}
        \SVM {The locally trained model}
        \SetKwFunction{SVM}{SVM}
        
        \Output{}     
        \MA {A matching array}        
% =========
$MA\gets []$\;
\ForEach(\tcp*[h]{A's rows}){$i \in D^A$}{% 
    \ForEach(\tcp*[h]{B's rows}){$j \in D^B$}{% 
        $MA$.\Append(\SVM{\Sim ($i$, $j$))}\;
    }
}
\Return MA\;
    \caption{Data Matching with Split-Learning SVM.}
    \label{algorithm:datamatching}
\end{algorithm}

\subsection{Synthetic Data Generation for Split Training}
In our Split Learning-based protocol, synthetic training data are individually generated at each party, utilizing its own recordset $r$, the common Reference Set, $RS$, and the attribute mapping $M$ between $r$ and $RS$. Algorithm \ref{algorithm:create_ts} illustrates this method. First, a corrupted version of $r$, $r'$ is built by performing Edit Distance operations (line 1). Then, a distance vector $D'$ is built for this corrupted dataset $r'$ through the data mapping process described above. For this corrupted version $D'$ and the original $D$, we will create a new feature vector by calculating the distances between the two vectors and label the pair $D_i$ and $D'_i$ with 1, as \textit{matching} (line 5). Then, to create non-matching pairs, records $k$ that are non matching with $i$ are selected from the same dataholder's recordset and the same operation is performed (line 6). For this approach to work properly, each party's recordset should have been deduplicated beforehand. Having created a synthetic training recordset, each party will now locally train its own model. Example \ref{ex:data_generation} clarifies this process.

\begin{Example}
\label{ex:data_generation}
Continuing the previous example, let's say that we have corrupted the record from Example \ref{ex:map_data_to_rs}, adding a character to the Surname field, and then we apply the same process. The result is a distance vector $D'$ with $D'=$ [[6, 3], [6, 6], [7, 2], [5, 5]]. Now, to create the feature vector, we need to compare $D$ and $D'$. To do so, we will employ cosine distance. As such, we have: $F$=[cos( [6, 3], [6, 3]), cos([5,5],[6,6]), cos([7,2], [7,2]), cos([5,5],[5,5])] = [0, 0.1414, 0, 0]. Since this vector represents a matching pair of records it gets labeled as matching or ``1''.
\end{Example}

\subsection{Split Data Matching}
After training concludes, matching parties exchange their Smashed Datasets which actually are distance vectors. Now, each party individually performs record matching, following the steps illustrated in Algorithm \ref{algorithm:datamatching}. For every item of the Distance Vector $D^A$ from Alice, its distance is calculated against every item of Bob's $D^B$, as during the training dataset generation, illustrated in Example \ref{ex:data_generation}. Then, the resulting vector is labeled as matching or non-matching using the locally trained SVM (line 4). The array of matches $MA$ is eventually produced (line 7). Finally, that each party delivers to the other the actual matching data.

\subsection{Discussion on Privacy Preservation}
Our Split Learning-based method has been designed to operate assuming a Honest But Curious setup. In this case, both parties try to infer as much information as possible adhering to the protocol. To prove the privacy-preserving characteristics of out method, we shall consider a series of attack types as described in \cite{vidanage2022taxonomy} and discuss how it performs in such situations. 

First, there are \textit{Dictionary Attacks}, where an adversary attempts to identify a sensitive value using a publicly available dictionary and encoding its values so as to match a dataset's encoded values. Our method is invulnerable to these attacks since each party receives from the other an array of distances (or similarities). As these functions are non-bijective (such a distance or similarity may be produced by more than one pairs of strings), the calculated distances between the dictionary entries and the RS cannot uniquely identify a real record.

Next, there are \textit{Frequency Analysis Attacks}, where an adversary uses a public dataset to study its distribution and identify quasi-identifier attributes. In our case, as the distances vector contains distances and these functions are non-bijective, there is no relation between the frequency of a dictionary entry and its distance from the Reference Set.

In \textit{Similarity Attacks}, an adversary may exploit the fact that distributions of similarities between encoded and plain text fields are maintained. so as to relate plain text values with encoded values. In our approach, Edit Distances are used, which are not limited by the bounds of similarity functions. As in previous cases, Edit Distance features non-bijectivity. On top of that, these distances are calculated against a custom made RS, unknown to the attacker. As such, launching a similarity attack is not feasible, as the attacker does not know either of the strings these distances refer to.

Finally, there are the \textit{Linkage and Ciphertext-only Attacks}, relying on linking publicly available information to reveal the quasi-identifiers. In Ciphertext-only attacks, the adversary analyses ciphertexts to recover plain texts. Let us consider the case where a dataholder tries to perform a brute-force attack. First of all, an external attacker will not know the Reference Set agreed by the matching parties. As such, and also considering that Edit Distance is non-bijective and that the intersection between the RS and the recordset is empty, even if an attacker manages to create a vector of distances identical to the recovered one, she has no evidence of the strings compared that will result in such a vector.

\begin{figure*}[t!]
\begin{minipage}[b]{0.47\textwidth}
         \includegraphics[width=\columnwidth]{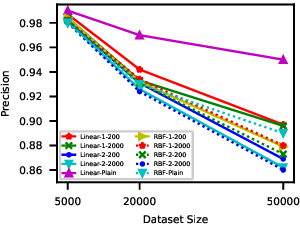}
          % \vspace{-0.15in}
        \subcaption{\footnotesize{Method Precision vs. RS size.}}
        \label{fig:reference_size_Precision}
    \end{minipage}
    % \hfill        
    \begin{minipage}[b]{0.47\textwidth}
        \includegraphics[width=\columnwidth]{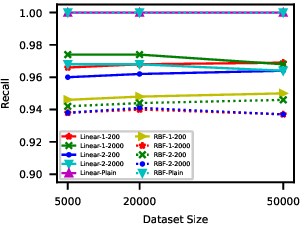}
          % \vspace{-0.15in}
        \subcaption{\footnotesize{Method Recall vs. RS size.}}
        \label{fig:reference_size_Recall}
    \end{minipage}
    \begin{minipage}[b]{0.47\textwidth}
            \includegraphics[width=\columnwidth]{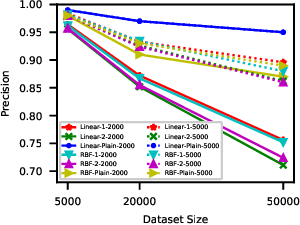}
          % \vspace{-.15in}
        \subcaption{\footnotesize{Method Precision vs. Training Set size.}}
        \label{fig:training_size_Precision}
    \end{minipage}
    % \hfill
    % \vspace{0.1in}
    \begin{minipage}[b]{0.47\textwidth}
        \includegraphics[width=\columnwidth]{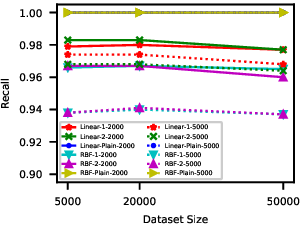}
          % \vspace{-0.15in}
        \subcaption{\footnotesize{Method Recall vs. Training Set size.}}
        \label{fig:training_size_Recall}
    \end{minipage}
    % \hfill
     \caption{Results of the Experimental Evaluation.}
    \label{fig:experiments}
\end{figure*}\section{Empirical Evaluation}
\label{sec:evaluation}
To evaluate the performance of our method, we conducted a series of experiments and measured matching quality and execution time. In this section, we present our results, but first we will describe the datasets we have used and our experimental setup.

\subsection{Experimental Setup and Datasets}
For our evaluation we have used Python 3, and scikit-learn\footnote{Implementation available at: \url{https://github.com/mikez3/Split-Learning-Privacy-Preserving-Record-Linkage}}. We ran our experiments on a Core i5-10600K host with 16GB of main memory powered by Ubuntu 22.04. 

We used recordsets coming from the North Carolina voters database\footnote{\url{https://dl.ncsbe.gov/?prefix=data}}, having used three samples with 5000, 20000 records and 50000 records each. In each sample, we have used the following string attributes: ‘last name’, ‘first name’ and ‘middle name’ and deduplicated the dataset so that these attributes form a candidate key. Then, we generated two databases. The first one belongs to Alice, while the second one belongs to Bob. Since we are interested in linking low quality data, we corrupted Bob’s records using the German Record Linkage Center’s data corrupter \cite{Corrupter}, with the dataset on Bob's side containing one error per row for the corrupted records, so that a join operation using these quasi-identifiers yields an empty result set. As our method features a Reference Set, we created two such Reference Sets of 200 and 2000 records using actor names from Wikipedia\footnote{\url{https://en.wikipedia.org/w/index.php?title=Category:20th-century_American_male_actors}}. These were built so that no attribute value from the recordsets is found in the Reference Sets.

%%%%%%

We measure matching performance, in terms of correctly matched record pairs using Precision and Recall. Precision is defined as the fraction of the relevant elements among the retrieved elements, while Recall is defined as the fraction of the retrieved relevant elements divided by the total relevant elements: $Precision = \frac{{TP}}{TP + FP}$ and $Recall = \frac{{TP}}{TP + FN}$. At this point, we have to stress that each party performs matching individually. As such, matching performance varies, depending on the performance of the SVM model trained locally. To this end, we report Precision and Recall separately so as to provide a more detailed view of our method's performance. We also report time efficiency in terms of testing time.

To have a broader view of our method, we considered two setups for our SVM classifier. The first setup employs a Linear SVM kernel with a hyperparameter C=100 at both parties, while the second one employs an RBF kernel with C=0.01 at both parties. We have used Edit Distance as a distance function between the Reference Set and each party's recordset and Cosine Distance to calculate the distances between the resulting distance vectors. For each setup, we have performed three executions and report the average results for Precision, Recall and matching time. Finally, to be able to have a basis of comparison, we also compare against a single local SVM classifier, as described in \cite{christen2008automatic}.

\subsection{Evaluation of Matching Performance}
To evaluate the matching performance and the behavior of our method, we will consider the size of the Reference Set used and the size of the Training Set used. We will vary these parameters and examine how our method behaves with respect to the dataset sample size used for matching in terms of Precision and Recall. 

\paragraph{Impact of Reference Set Size} Let us begin our assessment by varying the Reference Set size, while using a training dataset equal to 5000 records. The results of this set of experiments are illustrated in Fig. \ref{fig:reference_size_Precision} and \ref{fig:reference_size_Recall}. The horizontal axis represents the size of the dataset we attempted to match, while the vertical axis the respective measure, i.e. Precision or Recall. We report results both for the Linear and RBF kernels. For the results of the Split process, we illustrate results for both Reference Set sizes for each of the matching, designated by `-1-' for Alice and `-2-' for Bob, while the case of simple SVM, without privacy characteristics is designated by ``Plain''.

We will first present our Precision results (Fig. \ref{fig:reference_size_Precision}) starting with Plain SVM training, with Precision at 0.99 for the 5000 records datasets, dropping to 0.95 for the 50000 records datasets, indicating that Precision slightly drops as dataset size increases. Next, for the case of Split Learning, in all cases, Precision also drops, but faster, as the size of the matching datasets increases. RS size does not exhibit significant performance impact. However, there are slight fluctuations in Precision between Alice and Bob, something which is anticipated, since each of them has been trained using different data. This is the reason that Alice and Bob exhibit different performances using the Linear or the RBF kernel.

Recall results are illustrated in Fig. \ref{fig:reference_size_Recall}. First, we may discern that all Plain classifiers achieve absolute Recall, without being affected by matching set sizes. The performance of the Split Learning classifiers has not been deteriorated, in this case, as opposed to the case of Precision, when the size of the matching dataset increases. In this case, there is a drop in performance by approximately 3\% when Linear kernels have been used and up to 6\% when RBF kernels are used. For both parties the performance is aligned. On the other hand, the size of the Reference Set used does not incur any impact.

\paragraph{Impact of Training Set Size} To examine the impact of the Training Set size on matching performance, we change our setup keeping the RS size fixed to 2000 records, while we vary the size of the Training Set. Particularly, we consider two training setups, one with 2000 records and one with 5000 records. The results of this part of our evaluation for Precision and Recall are illustrated in Fig. \ref{fig:training_size_Precision} and Fig. \ref{fig:training_size_Recall} respectively. These figures have been set up in a similar way with  Fig. \ref{fig:reference_size_Precision} and Fig. \ref{fig:reference_size_Recall}. 

Starting with Precision, let us consider the case of Plain SVMs, for the Linear kernel. Precision starts at 0.99 for the 5000 records datasets, dropping to 0.95, when 50000 records are matched, regardless of the Training Set size. The Plain RBF kernel exhibits lower performance where performance drop is also steepest, with Precision values starting at 0.98 and dropping to 0.87. For Split Learning, Precision drops also, as the matching dataset size increases, but in a steeper manner. However, we may extract some additional useful conclusions. First of all, we may see that both Split Learning-based classifiers, `-1-' and `-2-' exhibit similar performance considering the same kernel and Training Set size. Next, it is easy to discern that, in terms of Precision, the larger Training Set (dotted lines) offers significantly better performance compared to the smaller one. This difference is significant, reaching 0.14 of Precision, for the case of the 50000 matching records. On the other hand, compared to the case of Plain models, using large Training Sets may deteriorate performance only by 10\%, which is the price to pay for protecting privacy. When matching smaller recordsets, this difference, in the case of 5000 records is only 1\%. Finally, using a smaller Training Set causes a more significant drop in performance, illustrated by a steepest slope in Fig. \ref{fig:training_size_Precision}, as the size of the matching datasets rise. This leads us to the conclusion that, when larger datasets are to be matched, larger Training Sets should be employed.

Recall results for this set of experiments are illustrated in Fig. \ref{fig:training_size_Recall}. Starting with the Plain SVM cases again, they achieve absolute Recall regardless of the kernel used. For Split Learning, the results are quite interesting, especially when considered in conjunction of those of Fig. \ref{fig:training_size_Precision}. To begin with, we may discern that Recall is not affected by the size of data to be matched, regardless the Training Set size or the kernel used for training. Nevertheless, using smaller Training Sets, designated by solid lines, results in higher Recall, approximately 0.98, compared to the cases of using the 5000 records datasets for training. However, in the latter case, Recall is between 0.96 and 0.99. This means a less than 4\% drop compared to the ideal case. Considering, now, the results for Precision as well, it is evident that using larger Training Sets results in high Precision at a very small cost of Recall, being a fair trade-off. To sum up, we have illustrated that at an expense of 10\% in Precision and less than 2\% in Recall, we have managed to provide privacy when matching 50000 records datasets. For smaller datasets (i.e. 5000 records), Precision may only drop by 0.01.

\begin{figure}[t]
    \centering
    \includegraphics[width=0.75\columnwidth]{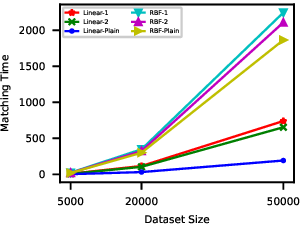}
    \caption{Matching times comparison.}
    \label{fig:matching times}
\end{figure}

\subsection{Evaluation of Time Performance}
Having seen how Split Learning-based Privacy-Preserving Matching behaves in terms of Precision and Recall, it would be interesting to see how much time these methods require to match data. First, of all, matching time is not affected by the size of the Reference Set used. To this end, we will report times for the cases we have used a Reference Set of 5000 records, with the cases of 2000 records exhibiting minimal fluctuations. These results are illustrated in Fig. \ref{fig:matching times}, with the horizontal axis representing the size of the dataset to be matched, while the vertical one stands for time in seconds.

Starting with Linear SVMs, for each of the two parties 8.6 and 6.7 secs are required, on average, to match 5000 records, respectively. For 20000 records, these times are 115.8 and 104.2 secs, while for 50000 records, 739.9 and 663.6 secs elapsed respectively. For the Plain SVM, these times are 2.4, 31.5 and 191.4 secs. It is evident that to offer privacy, almost four times the time of the Plain Linear SVM classifier is required. For the case of RBF SVMs, somewhat more time is needed to conclude the matching process. For 5000 records, 23 and 20 secs elapsed at each party on average. For 20000 records, 347.9 and 323.2 secs were required and, finally, matching 50000 records requires 2244.5 and 2107.6 seconds on average. For the Plain SVM classifier, these times are 19.5, 300.3 and 1864.1 secs, respectively.

%Record linkage is, by nature, a problem requiring quadratic time when blocking techniques \cite{Chr12} are not employed (a fact that also holds for Privacy-Preserving Record Linkage). This is somethiIn the case of the Linear kernel, the time required is approximately four times of that of the plain SVM case, while for the RBF case, a small constant overhead is incurred.

\section{Conclusions \& Future Work}
\label{sec:conclusions}
In this paper, a Split Learning-based Privacy-Preserving Record Matching using Support Vector machines and synthetic training data generation was presented. Its key characteristics are that it does not require a Linkage Unit, no data interchange takes place between data holders and that a small performance overhead is incurred, compared to the use of SVMs for plain text record linkage. Our future work is focused on improving the synthetic training set generation so as to further elevate matching results and to also apply differential privacy-based noise at the smashed-data, so as to also provide formal statistical privacy guarantees.

\bibliographystyle{ACM-Reference-Format}
\bibliography{sample-base}

% \appendix
\end{document}